\documentclass{article}
\usepackage{ijcai26}

\usepackage{times}
\usepackage{soul}
\usepackage{url}
\usepackage[hidelinks]{hyperref}
\usepackage[utf8]{inputenc}
\usepackage[small]{caption}
\usepackage{graphicx}
\usepackage{amsmath}
\usepackage{amsthm}
\usepackage{booktabs}
\usepackage{algorithm}
\usepackage{algorithmic}
\usepackage[switch]{lineno}

\title{Invisible Agents, Uninformed Patients: Towards Responsible Deployment Of Autonomous AI Diagnostic Agents In Sub-Saharan Africa}

\author{
Percy Brown$^1$
\and
Kweku Yamoah$^2$
\affiliations
$^1$Independent Researcher \\%
$^2$University of Florida, Gainesville, FL, 32611 \\%
}

\begin{document}

\maketitle

\begin{abstract}
Autonomous AI diagnostic agents, systems that analyse patient-specific clinical data and produce diagnostic outputs or triage decisions without mandatory real-time human review, are increasingly deployed across eHealth platforms in sub-Saharan Africa at a pace that has outrun the governance infrastructure needed to oversee them. While significant bodies of work address AI accountability, transparency and explainability in healthcare, existing frameworks are largely clinician-centered and assume regulatory conditions that do not uniformly exist in low-resource settings. A patient-centered analysis of the disparity in patient awareness regarding autonomous agents, which results in a structural accountability gap, is mostly missing from the literature. This paper synthesizes existing research on informed consent, algorithmic accountability, and explainable AI to highlight three distinct challenges introduced by deploying AI agents in the sub-Saharan African context. Drawing on three documented deployment cases, including computer-aided tuberculosis detection in Tanzania, diabetic retinopathy and TB screening in Zambia, and mobile health chat-bot triage in Ghana, it demonstrates that these gaps are already present in active deployments across the region. In response, the paper proposes three foundational principles; agent-aware informed consent, human override as a structural requirement and contextually adapted explainability. This triad of principles lays a practical minimum standard for developers, health system administrators and policymakers in contexts where formal AI regulation remains nascent.
\end{abstract}

\section{Introduction}
\label{s1}

The deployment of AI in healthcare in sub-Saharan Africa has increased in recent years, spanning tuberculosis screening, diabetic retinopathy detection, maternal health assessment, and mental health diagnosis \cite{ephraim2024,njei2023}. Facing persistent shortages of specialist clinicians and laboratory infrastructure, sub-Saharan Africa presents a compelling case for AI-assisted diagnostics as a means of extending reach without proportional increases in human expertise.

However, not all AI tools used in healthcare carry the same implications for patients. A decision-support tool that provides information for a clinician to interpret differs from an autonomous AI agent that analyzes patient data and produces a diagnostic output or a triage decision with minimal or no real-time human involvement. The latter raises important questions. Does the patient know that an autonomous agent is acting on them and, if not, are their rights any less real?

Research in first-world countries has established that autonomous AI diagnostic systems present accountability and transparency challenges that conventional medical liability and informed consent frameworks were not designed to handle \cite{habli2020,kiseleva2022}. In sub-Saharan Africa, where regulatory infrastructure is thinner and patients range from digitally literate to those with minimal exposure to AI-enabled services, these challenges take a more urgent form.

Existing literature on AI accountability and explainability in healthcare is substantial but has three notable limitations in this context. First, it is predominantly clinician-centred rather than patient-centred \cite{obaid2025}. Second, it assumes regulatory and institutional conditions such as designated national AI authorities that do not uniformly exist in sub-Saharan Africa. Third, it has not addressed the particular challenge of autonomous agents operating in low-literacy, low-regulation, and mobile-first health contexts. This paper argues that autonomous AI diagnostic agents represent a categorically distinct accountability challenge that requires principles designed specifically for the sub-Saharan context. 

The remainder of this paper is structured as follows: Section~\ref{s2} reviews related work and identifies gaps, Section~\ref{s3} examines accountability challenges with case studies, Section~\ref{s4} presents the proposed framework,  Section~\ref{s5} discusses stakeholder implications, Section~\ref{s6} outlines future work, and Section~\ref{s7} concludes.

\section{Background and Related Literature}
\label{s2}
This section situates the study within the existing body of research on AI diagnostic systems and their governance. It first examines the emergence and application of AI diagnostic agents in sub-Saharan Africa, highlighting both their potential and the pace of their deployment. It then reviews key debates on accountability, transparency, and informed consent in health AI, with particular attention to their underlying assumptions. Finally, it identifies critical gaps in the literature, including the limited focus on patient awareness, the misalignment of existing governance frameworks with sub-Saharan African contexts, and the absence of empirical evidence on how patients understand and interact with autonomous diagnostic systems.

\subsection{AI Diagnostic Agents in Sub-Saharan Africa}
\label{s2a}

Sub-Saharan Africa carries a disproportionate share of the global burden of communicable and non-communicable diseases while facing persistent shortages of trained clinicians, laboratory infrastructure and specialist expertise \cite{ceimia2024,oladipo2024}. As a result, AI diagnostic tools offer the prospect of extending screening and triage capacity into underserved settings, improving early detection rates and reducing the burden on overstretched health workers.

Documented applications across the region include AI-powered diabetic retinopathy screening in Zambia, deep learning tools for retinal disease detection in East Africa, AI-assisted pathology for cervical cancer screening in West Africa, and multi-agent systems for clinical triage and patient communication in Ghana and East Africa \cite{bellemo2019,ephraim2024}. The pace of deployment is accelerating, with mobile health platforms beginning to integrate AI-driven triage functions accessible directly to patients without clinical intermediaries, thereby producing outputs acted upon without meaningful human review due to workflow pressures, infrastructure constraints or deliberate design choices.

\subsection{Accountability and Transparency in Health AI}
\label{s2b}

A substantial literature addresses AI accountability and transparency in healthcare. Recurring themes include the opacity of black-box models, the difficulty of assigning liability for AI-driven errors, the erosion of informed consent, and the risk of algorithmic bias amplifying existing health inequities \cite{mittelstadt2016,obermeyer2019,habli2020}. These concerns are increasingly reflected in instruments such as the EU AI Act and WHO guidance on AI in health.

Research on explainability has grown substantially, with \cite{kiseleva2022} identifying transparency as a multilayered system of accountabilities spanning technical interpretability, organisational disclosure, and legal compliance. \cite{wachter2017} and \cite{doshivelez2017} have similarly argued that explainability is not a single property but a set of audience-specific requirements. These contributions are foundational, but they are primarily oriented toward clinician and regulator audiences in high-income settings.

On informed consent specifically, substantial work has examined how autonomous systems challenge traditional consent frameworks. \cite{mittelstadt2016} argued that algorithmic decision-making introduces accountability gaps that existing ethical and legal frameworks were not designed to address. \cite{char2018} examined how AI tools alter the physician-patient relationship and complicate disclosure obligations. However, these analyses centre on health systems with established clinical governance and patient advocacy infrastructure, which are conditions that differ from much of sub-Saharan Africa.

In the African context, \cite{ibeneme2021} emphasised the importance of inclusive AI governance frameworks that account for Africa's unique socioeconomic conditions, while \cite{audencio2025} argued that the global AI governance agenda risks producing inequitable outcomes for health in sub-Saharan Africa by failing to centre affected populations. \cite{path2024} cautioned explicitly that the field is not yet ready for fully autonomous diagnostic applications without robust oversight mechanisms.

\subsection{Gaps in the Existing Literature}
\label{s2c}
Three gaps motivate this paper. First, the literature is predominantly focused on clinician perspectives and system-level accountability, with limited attention to patient awareness of autonomous agent involvement as a distinct governance problem. As \cite{obaid2025} acknowledged in their qualitative study of UK healthcare professionals, research has not yet adequately explored how AI-assisted care affects patient autonomy, trust, and understanding from the patient's own standpoint. This reflects how the field has been framed from the outset, with clinician experience treated as the primary lens through which AI accountability is understood.

Second, existing governance frameworks assume regulatory and institutional conditions that do not uniformly exist in sub-Saharan Africa. The EU AI Act, for instance, presupposes national authorities with technical capacity to audit AI systems \cite{euaiact2024}. As of 2024, only seven African nations had drafted national AI strategies, and none had implemented formal AI regulation \cite{brookings2024}. The AU Continental AI Strategy, endorsed in July 2024, is currently in its first implementation phase (2025–2026), which focuses on establishing governance structures and mobilising resources \cite{whitecase2024}, meaning that enforceable AI governance mechanisms do not yet exist across the region even under this framework. Therefore, most African countries lack dedicated AI regulatory bodies and operate health AI deployments under frameworks designed for conventional medical devices or general data protection that are not adequate for autonomous diagnostic agents \cite{ibeneme2021}.

\begin{figure*}[!htbp]
    \centering
    \includegraphics[height=0.5\linewidth, keepaspectratio]{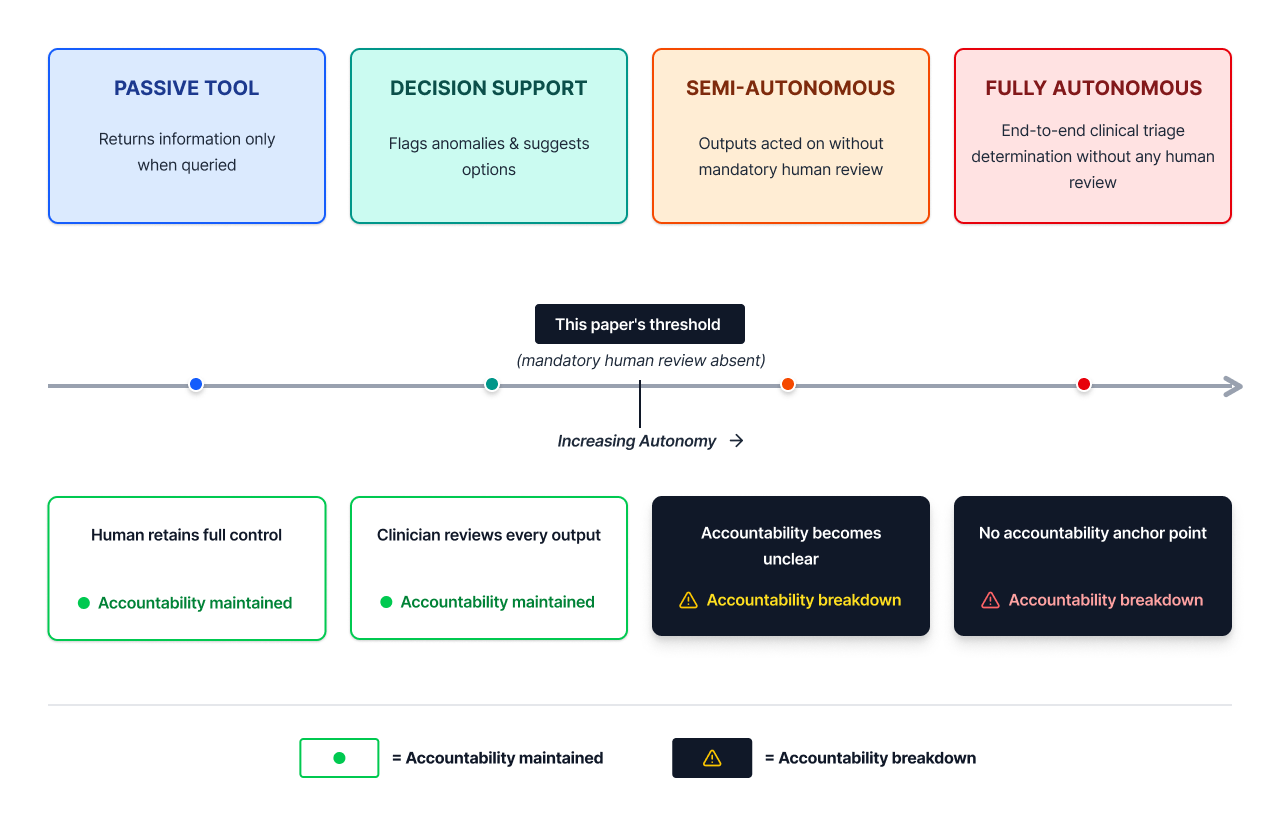}
    \caption{The Autonomy Spectrum in AI Diagnostic Systems}
    \label{fig1:placeholder}
\end{figure*}

Third, no systematic empirical analysis exists of how patients in sub-Saharan African eHealth contexts experience or understand autonomous agent decision-making. This absence is foundational, because every accountability mechanism premised on patient awareness presupposes a level and distribution of that awareness which has never been established in this context. The conditions that mostly shape patient understanding in sub-Saharan Africa, such as uneven digital literacy, mobile-first health access, linguistic diversity, and limited prior exposure to AI-enabled services are materially different from those studied elsewhere, therefore, findings from high-income contexts cannot be assumed to transfer. This paper addresses that gap conceptually and empirical investigation remains an urgent priority for future research. 

\section{Analysis: The Distinct Accountability Challenge}
\label{s3}
This section develops the core analytical argument of the paper by examining how accountability challenges emerge in the deployment of autonomous AI diagnostic agents. It begins by defining such agents along a spectrum of autonomy to clarify where accountability shifts occur. It then introduces the concept of a two-level transparency deficit, distinguishing between technical opacity and failures in patient-facing communication. Building on this, it analyses the implications for informed consent in contexts where patient awareness of AI involvement cannot be assumed. Finally, it grounds the analysis in illustrative cases from sub-Saharan Africa, demonstrating how these challenges manifest in real-world deployments and why existing governance mechanisms remain insufficient.

\subsection{Defining Autonomous AI Diagnostic Agents}
\label{s3a}

The term \textit{autonomous AI diagnostic agent}requires precise definition, given that real-world systems operate on a spectrum rather than a binary, as illustrated in Figure~\ref{fig1:placeholder}. Scholars in AI and health informatics have distinguished between tools that augment human decision-making and those that replace it \cite{russell2020,topol2019}. At one end of the spectrum sits a purely passive tool: a database or reference system that returns information only when queried by a clinician. Moving along the spectrum, decision-support tools actively surface recommendations or flag anomalies, but a human clinician retains interpretive authority and clinical responsibility. Further along, semi-autonomous systems produce outputs that may be acted upon without mandatory human review, typically under workflow or infrastructure pressure. At the fully autonomous end, systems analyse patient data and produce diagnostic outputs or triage decisions that are treated as final determinations without any human review step.

\begin{figure*}[!htbp]
    \centering
    \includegraphics[height=0.5\linewidth, keepaspectratio]{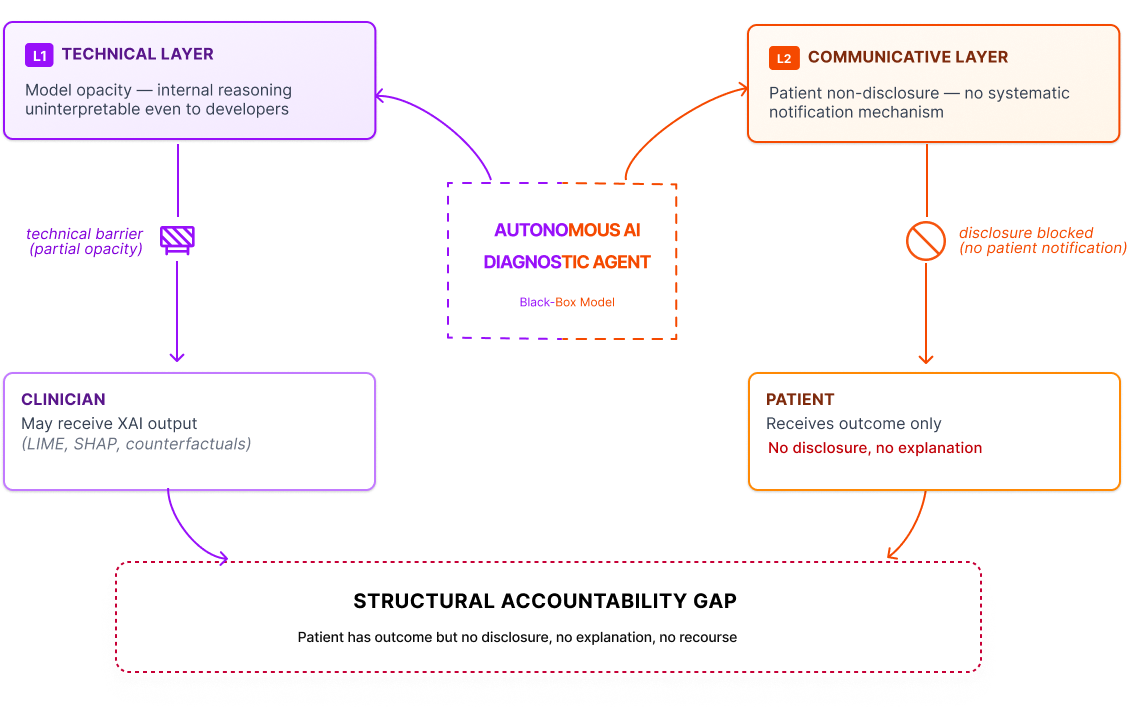}
    \caption{The Two-Level Transparency Deficit in Autonomous AI Diagnostic Agents}
    \label{fig2:placeholder}
\end{figure*}

For the purposes of this paper, an autonomous AI diagnostic agent is defined as a system that meets three conditions: 
\begin{enumerate}
    \item Analyses patient-specific clinical data such as symptoms, images or test results without requiring real-time clinician input
    \item Produces an output, such as a diagnosis, triage classification, or clinical recommendation, that is intended to be acted upon and
    \item No mandatory human review step exists between the system's output and the resulting clinical action.
\end{enumerate}
This definition deliberately captures systems operating at the semi-autonomous and fully autonomous end of the spectrum, where the accountability implications are qualitatively different from those of decision-support tools \cite{coiera2019,habli2020}. It excludes systems where a qualified clinician reviews and takes responsibility for every output before any clinical action is taken.

This distinction matters because accountability usually follows the decision-maker. Where a human clinician reviews and endorses an AI output, accountability remains with that clinician. Where no such review occurs by design or by default, accountability becomes structurally unclear, and the patient is left without recourse.

\subsection{The Two-Level Transparency Deficit}
\label{s3b}

The accountability challenge is compounded by a transparency deficit operating at two distinct levels, as illustrated in Figure~\ref{fig2:placeholder}. The first level is technical, where many diagnostic AI systems, especially those using deep learning, produce outputs whose internal reasoning is uninterpretable even to their developers \cite{kiseleva2022,mittelstadt2016}. XAI research has made progress on post-hoc explanation methods, but these remain largely oriented toward technical audiences rather than patients.

The second level is communicative, where patients are rarely informed in any meaningful way that an autonomous system is involved in their diagnostic pathway, what the system does, or what rights they hold in relation to its outputs. In sub-Saharan Africa, some patients with higher digital literacy or urban access to technology may recognise or suspect autonomous agent involvement. Others, especially in rural areas or with lower formal education, may have no such awareness. The structural problem is not simply that some patients are unaware. It is that whether a patient happens to be aware currently determines whether they receive adequate protection. No systematic mechanism guarantees equitable disclosure.


\subsection{The Consent Dimension}
\label{s3c}

Informed consent in medicine requires that patients understand what is being done to them, including by what means, which is a foundational condition for the exercise of patient agency. When an autonomous agent analyses a patient's data and produces a clinical output, that constitutes a medical act. The patient's right to know it is happening, to refuse it, or to seek human review is not a novel demand but an extension of established consent principles \cite{char2018}.

Yet as \cite{kiseleva2022} argued, compliance with informed consent requirements in the AI context depends not only on clinical knowledge but on comprehension of AI systems. This cannot be assumed in any patient population and is more uncertain in sub-Saharan Africa where formal AI literacy is minimal and systems are sometimes accessed directly through mobile interfaces with no disclosure mechanisms built in. 

The convergence of these three factors, i.e. the broken accountability chain, the two-level transparency deficit, and the inadequacy of existing consent frameworks constitutes the core challenge this paper responds to.

\subsection{Illustrative Cases from Sub-Saharan Africa}
\label{s3d}

The challenge described in Sections~\ref{s3a} through~\ref{s3c} is not theoretical. Three documented deployment cases from sub-Saharan Africa demonstrate what each component of the gap looks like in practice. Taken together, they illustrate how the broken accountability chain, two-level transparency deficit, and consent inadequacy manifest simultaneously in real deployments, and why none of the existing governance mechanisms reviewed in Section~\ref{s2} are sufficient to address them.

\subsubsection{Computer-Aided Tuberculosis Detection in Tanzania}
Computer-aided detection (CAD) software systems such as CAD4TB has been deployed for tuberculosis screening across multiple sub-Saharan African countries including Tanzania, where the country carries one of the highest TB burdens globally. \cite{breuninger2014} conducted one of the earliest validations of CAD4TB in two cohort studies in Tanzania, finding that the software accurately distinguished between culture-positive TB cases and controls with an area under the ROC curve of 0.84. Critically, the authors noted that further studies on operational and ethical aspects were needed to determine the system's appropriate place in diagnostic algorithms, a recommendation that has not been uniformly acted upon in subsequent deployments.

The operational reality in Tanzania and comparable settings is that CAD systems are deliberately designed to function without trained human readers present, and this is explicitly their primary value proposition in settings where radiologists are scarce \cite{codlin2023}. The Stop TB Partnership's practical guide for CAD deployment documents that these systems are intended to replace human interpretation in resource-constrained contexts, and that the CAD4TB system can be deployed in remote settings without internet access and without trained human readers on site. In practice, this means that patients in TB screening programmes in Tanzania may receive a classification about whether they have tuberculosis by an autonomous system, without a radiologist or clinical officer reviewing the output before any follow-up action is initiated.

What is absent from this deployment model is any patient-facing accountability mechanism. The patients receiving these classifications are not told that an AI system has analysed their X-ray, are not given a plain-language explanation of how the determination was made or what its error rate is, and have no formal pathway to request that a human clinician verify the output before they are sent for further testing or, in a false negative scenario, discharged. The TB screening context is consequential in this regard: a false negative means a patient with active tuberculosis is not identified and treated, with implications both for the individual's health outcome and for community transmission. The accountability chain is severed by design, and none of the existing governance instruments in Tanzania such as the Tanzania National AI Strategy Framework \cite{tanzania_ai_strategy2025}, provide any mechanism to close this gap.

\subsubsection{Diabetic Retinopathy and TB Screening in Zambia}

\cite{bellemo2019} conducted a clinical validation study of a deep learning system for diabetic retinopathy screening in Zambia, which was one of the first such studies in a sub-Saharan African context. The study found that the AI system achieved clinically acceptable performance for identifying referable retinopathy. The study was subsequently cited as evidence for the feasibility of AI-assisted retinopathy screening across the region, and the system was deployed beyond the research context into clinical use. A more recent multi-site validation study conducted at three health facilities in Zambia between 2021 and 2023 further confirmed that cloud-based AI systems for chest X-ray analysis could match the performance of radiologists in high TB-burden settings \cite{kazemzadeh2024}.

What neither the validation studies nor subsequent deployment documentation addresses is the patient experience of these systems. In both the retinopathy and TB contexts, patients attending health facilities in Zambia had no formal mechanism to understand that an automated system was producing the clinical determination about their condition, request that a human clinician verify the AI output before clinical action was taken, or receive an explanation of how the determination was made and what its limitations were. The validation literature establishes diagnostic accuracy. It does not establish, and in most cases does not address, whether patients were informed, whether they consented to autonomous AI involvement in their diagnosis, or whether they had any recourse if the system produced an incorrect or ambiguous result. Like Tanzania, the Zambian AI Strategy 2024-2026 \cite{zambia_ai_strategy2024} does not provide any mechanism to close this gap. This is the communicative dimension of the transparency deficit. Although the technical performance of these systems may be adequate, the patient's relationship to the decision-making process remains unaddressed.

\subsubsection{Mobile Health Chatbot Triage in Ghana}

Ghana represents a case where the accountability gap extends beyond hospital-based diagnostic tools to direct-to-patient AI interfaces accessed entirely through mobile telecommunications infrastructure, systems with which patients interact without any clinical intermediary present. Several major mobile network operators and digital health platforms operating in Ghana have deployed AI-enabled chatbot systems that collect patient symptoms through conversational interfaces, apply algorithmic processing, and produce health recommendations, risk classifications, or referral decisions \cite{dzando2022}. Under the definition established in Section~\ref{s3a}, these constitute autonomous diagnostic or triage outputs. They analyse patient-specific data, produce actionable clinical recommendations, and do so without any mandatory human clinician review step in the patient-facing interaction.

The scale of this deployment mode is significant. Mobile-first health services are among the fastest-growing points of healthcare contact for urban and peri-urban populations who lack convenient access to formal health facilities. This means that the chatbot interface is not a supplement to a clinical encounter. For some patients, it may be the primary point of triage, and the AI's recommendation is the primary basis on which they decide whether to seek in-person care. In this context, the communicative transparency deficit identified in Section~\ref{s3b} takes on particular urgency. A patient who interacts with a symptom-checking chatbot on their mobile phone may have no way of knowing, from the interface itself, that an AI system is making a health determination about them, how confident the system is in that determination, what the system does not know, or how to challenge the output.

 In the meantime, Ghana has taken some steps such as launching a National AI Strategy (2023-2033) \cite{ghana_ai_strategy2023} and an AI Practitioners' Guide \cite{heritors2025gaipg}, but these instruments are strategic and advisory rather than enforceable. They do not yet translate into requirements that AI health system developers and operators must disclose agent involvement to patients, provide override pathways, or explain algorithmic outputs in accessible forms.

\subsubsection{The Common Pattern}
Taken together, the three cases examined above reveal a consistent structural pattern rather than three isolated incidents. In Tanzania, the accountability gap is built into the clinical deployment model itself. CAD systems are designed by specification to replace human readers, which means the absence of patient disclosure and override pathways is a design choice, not an oversight. In Zambia, the gap emerges in the transition from validated research tool to deployed clinical system. The rigour applied to technical performance in the validation phase does not carry over into patient-facing accountability mechanisms in the deployment phase. In Ghana, the gap takes the form of a specific regulatory vacuum, with mobile health platforms operating at scale without any enforceable requirement to disclose AI agent involvement, provide override mechanisms, or explain algorithmic outputs in accessible terms.

What unites all three cases is that autonomous AI diagnostic outputs are being acted upon without patient disclosure, without mandatory human review pathways, and without accessible explanations. In each case, the existing governance instruments in place are insufficient to require otherwise. This is not primarily a technical problem. The systems involved have demonstrated acceptable diagnostic performance in their respective validation studies. The gap is in governance and design, where the principles, requirements, and structural safeguards that would translate technical capability into accountable clinical practice have not been established. Better model accuracy will not close a gap that is defined by the absence of disclosure, override, and explainability requirements. That requires the structural response the three principles in Section~\ref{s4} are intended to provide.

\begin{figure*}[!htbp]
    \centering
    \includegraphics[height=0.5\linewidth, keepaspectratio]{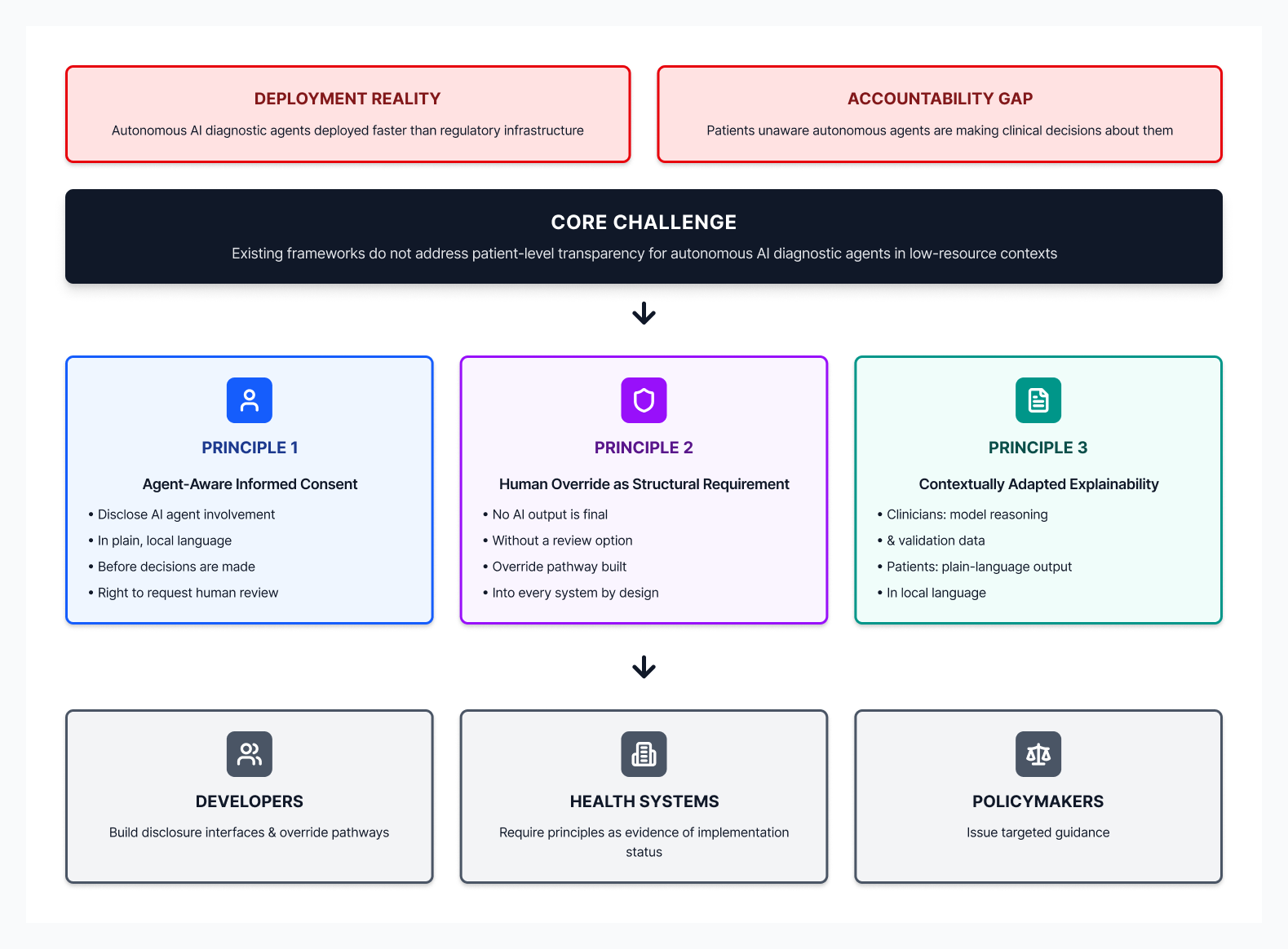}
    \caption{Conceptual framework for responsible deployment of autonomous AI diagnostic agents in sub-Saharan Africa}
    \label{fig3:placeholder}
\end{figure*}

\section{Proposed Framework}
\label{s4}

The three principles proposed in this section are derived through direct mapping from the three components of the accountability challenge identified in Section~\ref{s3}. 
\begin{enumerate}
    \item The consent dimension (Section~\ref{s3c}) motivates Principle 1. If patients are entitled to informed consent and AI involvement undermines that entitlement, then the response must guarantee disclosure at the patient level.
    \item The broken accountability chain (Section~\ref{s3a}) motivates Principle 2. If autonomous outputs have no mandatory human review, the structural response is to require one.
    \item The two-level transparency deficit (Section~\ref{s3b}) motivates Principle 3. If the reasoning behind autonomous outputs is neither technically interpretable nor communicated to affected patients, the response must address both dimensions in a contextually appropriate form.
\end{enumerate}
Taken together, the three principles constitute a minimum viable framework. They do not resolve every governance question, but they directly address each component of the identified challenge. Figure~\ref{fig3:placeholder} presents the conceptual framework illustrating how the principles respond to the core challenge and translate into stakeholder-specific obligations.

\subsection{Principle 1: Agent-Aware Informed Consent}
\label{s4a}
When an autonomous AI agent is involved in making a diagnostic or triage determination, patients should be informed in plain language accessible at their literacy level and in their primary language. This principle applies regardless of whether individual patients may already suspect agent involvement because incidental awareness is not equivalent to guaranteed disclosure. The disclosure should make clear that a computer system rather than a human clinician has analysed their data, that the system makes decisions based on patterns learned from data rather than individual clinical judgement, and that patients have the right to request human review of the output.

This principle extends rather than replaces existing informed consent doctrine. \cite{char2018} and \cite{mittelstadt2016} have argued that AI systems require consent frameworks to evolve beyond disclosure of treatment risks toward disclosure of the nature of the decision-making process itself. This principle gives that argument practical form in a low-resource, mobile-first context. Disclosure mechanisms must be designed for accessibility rather than legal compliance. Verbal notification by a health worker, a brief visual indicator on a mobile interface, or a standard statement in a local language at point of contact are all preferable to a lengthy technical consent form that achieves disclosure in name only.

\subsection{Principle 2: Human Override as a Structural Requirement}
\label{s4b}
No autonomous AI diagnostic agent should be permitted to produce a clinical outcome that is final and unreviewable. Every system should be designed with a human override pathway, a mechanism by which a patient or clinician can request that the autonomous output be reviewed by a human professional before clinical action is taken. This should not be a luxury reserved for well-resourced systems, but a fundamental patient safety requirement.

This principle draws on and extends calls in the literature for meaningful human control over consequential AI decisions \cite{habli2020,path2024}. Where existing literature frames human oversight as a desirable property of AI system design, this principle frames it as a non-negotiable structural requirement of deployment, and one that must be verified at the procurement stage. System design, workflow integration, and deployment contracts with health institutions should all be evaluated against a single question: can a patient or clinician stop this agent's output from taking clinical effect without human review? If the answer is no, the deployment does not meet a basic standard of responsible practice, regardless of the system's technical performance.

\subsection{Principle 3: Contextually Adapted Explainability}
\label{s4c}
Transparency requires that the reasoning behind an autonomous agent's output be communicable to those it affects. The XAI literature has established a rich set of methods for generating post-hoc explanations of model outputs including LIME, SHAP, and counterfactual explanations. However, these have been designed primarily for technical and clinical audiences \cite{arrieta2020}. In sub-Saharan Africa, with direct-to-patient AI interfaces growing through mobile health platforms, the explainability challenge must extend to patients themselves.

Contextually adapted explainability means the explanation of an AI agent's output should be calibrated to the audience receiving it. For clinicians, this means model-level interpretability tools and documentation of training data provenance, validation performance on locally representative populations, and known limitations, which are consistent with calls for algorithmic transparency and bias documentation \cite{obermeyer2019}. For patients, it means a plain-language account of what the system found and why, without requiring technical literacy, in a form accessible in the relevant local language. This may be demanding but not impossible, and it is a condition of genuine transparency.

\section{Discussion}
\label{s5}
This section interprets the implications of the proposed framework for key stakeholders involved in the design, deployment and governance of autonomous AI diagnostic agents in sub-Saharan Africa. It translates the three principles into actionable considerations for developers, health system institutions, and policymakers. In doing so, it highlights how responsible deployment can be advanced even in the absence of comprehensive regulatory frameworks, by embedding accountability mechanisms directly into system design, procurement practices, and policy guidance.

\subsection{Implications for Developers}
For developers of AI diagnostic agents intended for deployment in sub-Saharan Africa, the three principles translate into concrete design and documentation requirements. Disclosure interfaces should be built in as standard features, not compliance afterthoughts. Human override pathways should be architected into clinical workflows at the design stage. Explainability modules should be developed in parallel with diagnostic functionality, and localised language versions should be a deployment prerequisite rather than a post-launch enhancement. Developers should also document validation data from populations comparable to the intended deployment context, given the well-established risk that models trained on non-representative data reproduce or amplify existing health inequities \cite{obermeyer2019,audencio2025}.

\subsection{Implications for Health Systems/Institutions}
For health system administrators and ministries of health, the principles provide a basis for procurement standards and deployment agreements. Before integrating an autonomous AI diagnostic agent into a health facility's workflow, administrators should require documented evidence of agent-aware disclosure mechanisms, human override pathways, and performance validation on locally representative populations. The absence of a tailored regulation does not prevent health institutions from establishing their own standards of responsible practice. In contexts where regulatory frameworks are still growing, institutional procurement standards represent the most immediately actionable governance lever.

\subsection{Implications for Policymakers}
For policymakers across sub-Saharan Africa, this paper makes the case that AI governance in health cannot wait for comprehensive AI legislation. Targeted guidance focused specifically on autonomous AI agents in clinical pathways and requiring the three principles as a baseline would address the most urgent risks without requiring full regulatory infrastructure to be in place first. This is consistent with international calls for context-sensitive and incremental AI governance in low-income settings \cite{ibeneme2021,ceimia2024}.

\begin{figure*}[!htbp]
    \centering
    \includegraphics[height=0.5\linewidth, keepaspectratio]{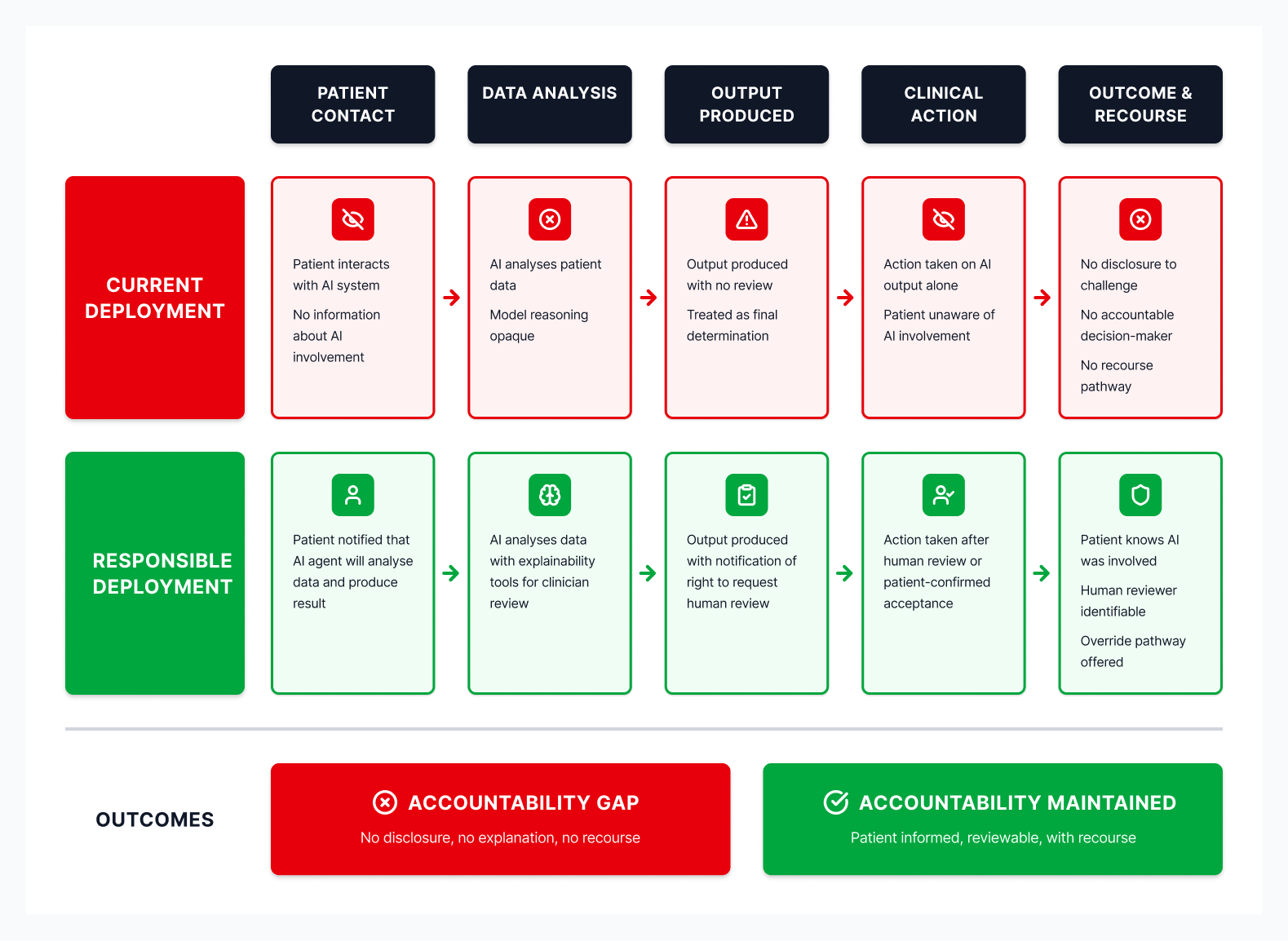}
    \caption{Patient journey under current autonomous AI deployment versus responsible deployment applying the three proposed principles across five stages of a diagnostic encounter}
    \label{fig4:placeholder}
\end{figure*}

\medskip
\noindent Figure~\ref{fig4:placeholder} illustrates the practical difference the three principles make at each stage of a patient's diagnostic encounter. The left column traces what currently happens in deployments where no disclosure, override, or explainability requirement is in place, where the patient receives a clinical outcome with no knowledge of AI involvement and no recourse if it is wrong. The right column shows how the same five stages would unfold if the three principles were applied by design, demonstrating that the governance requirements proposed in this paper are not abstract obligations but concrete changes to system architecture and clinical workflow.

\section{Future Research}
\label{s6}
The three principles are proposed not as evidence-based conclusions but as theoretically grounded hypotheses derived from synthesis of the accountability, consent and explainability literature, mapped against documented deployment realities in sub-Saharan Africa. As such, the absence of primary empirical data reflects a deliberate feature of the contribution's scope rather than a gap, and the principles are proposed as a framework for future empirical investigation.

Therefore, future work should investigate the extent to which patients in the region currently understand when autonomous agents are making decisions about them, audit additional existing deployments against the three principles proposed here, test disclosure instruments across varied literacy and language contexts, and examine the feasibility of human override pathways in resource-constrained health facilities. The illustrative cases presented in Section~\ref{s3d} provide a starting point for such investigations.

\section{Conclusion}
\label{s7}
Autonomous AI diagnostic agents hold genuine promise for extending healthcare reach in sub-Saharan Africa, but this promise must not come at the expense of accountability. The enthusiasm for what AI can do in health in Africa does not resolve the question of who is accountable when it acts autonomously and gets something wrong. This paper has argued that autonomous agents represent a categorically distinct accountability challenge from conventional decision-support tools, and one that existing African governance frameworks are not currently equipped to address.

The three principles do not resolve every question this field will need to answer, but they address the most immediate gap, which is ensuring that patients, whether already aware of agent involvement or not, have a structural guarantee of disclosure, the ability to seek human review, and access to explanations they can actually use. 

\newpage
\bibliographystyle{named}
\bibliography{ijcai26}

\end{document}